# The Effect of the 12μm Band: Comparing GOES-11 and GOES-12 Data Using the 3-Channel Volcanic Ash Algorithm

Emily Matson

**Student:** Matson, Emily M.

**Firms:** Center for Satellite Applications and Research (NOAA/NESDIS)

**Mentor:** Kenneth L. Pryor

**Project Title:** The Effect of the 12μm Band: Comparing GOES-11 and GOES-12 Data Using the 3-Channel Volcanic Ash Algorithm

*Background*

From April 2003, the split window 12μm channel has not been available on the GOES imagers. It will not be available until about 2013. By hovering over the same place on the Earth's surface, the GOES (Geostationary Operational Environmental Satellite) satellites provide the continuous data needed for concentrated data analysis. The 12μm channel, used in the GOES-11 (or GOES-West) Imager has been replaced by a 13.3μm channel in the GOES-12 (or GOES-East) Imager. There has been concern that the 13.3μm channel will not be as accurate in collecting volcanic ash data, especially when there is a significant amount of high cirrus cover or diffuse ash. This is of great importance because volcanic ash clouds, if undetected, can be dangerous and even fatal for planes. If the ash is ingested into the jet engines, volcanic ash decreases their efficiency. Furthermore, the plane's leading edge surfaces can be damaged. Recently, the increased amount of air traffic in the circumpacific region of the globe where the most potentially active volcanoes are located has increased the need for accurate volcanic ash data. There is also expected to be degradation in the detection of low clouds. Differences in cloud data have important implications for future weather forecasting and interpretation.

*Description*

This paper provides a comparison of various types of cloud data as detected in infrared (IR) images for GOES-11 and GOES-12 for the Mexican volcanoes Colima and Popocatepetl. No volcanic ash data was available due to poor data and lack of activity. These volcanoes are particularly useful because they are monitored by both GOES-11 in the west and GOES-12 in the east (since there is an overlap). Both satellite imagers use a specific algorithm designed to highlight volcanic ash in the IR images. Each algorithm is an arithmetic combination of the brightness temperatures as measured by three channels: one longwave, one window, and one shortwave. The result, which is displayed in the IR images, is a brightness count (unitless) in which different numerical values are assigned different shades of gray. The major difference is that the GOES-11 algorithm utilizes Channel 5, the 12μm channel, whereas the GOES-12 algorithm utilizes Channel 6, the 13.3μm channel. The IR images from GOES-11 and GOES-12 are compared to visible (VIS) images generated by GOES-12. In addition, scatter plots of brightness temperature vs. brightness count are used for both satellite imagers to help determine patterns and data correlations. The data indicates that, in general, the GOES-11 volcanic ash algorithm with the 12μm band provides much better data for various types of clouds. It emphasizes the importance of the 12μm band and provides scope for future research in this area.

1. Introduction

In April 2003 Geostationary Operational Environmental Satellite (GOES)-12 was implemented as the primary spacecraft to monitor weather conditions over North and South America for GOES-East. As a geosynchronous satellite, GOES-12 has the same period of revolution as the Earth, and is thus always able to hover above the same point on the Earth's surface, thus providing the continuous data needed for concentrated data analysis. Previous GOES satellites, including GOES-11, had utilized the 12μm "split window" band (Channel 5). However, a 13.3μm band has replaced the 12μm band for GOES-12. This was due to the need for the 13.3μm band for more accurate cloud top height estimates using a technique called $CO_2$ slicing. The 12μm band was originally to be included along with the 13.3μm band in GOES-12, but insufficient budgeting prevented this (Ellrod, 04a).

In the GOES-11 volcanic ash algorithm, previously used with GOES-East, brightness temperatures were taken for Channels 2 (3.9μm: shortwave infrared (IR)), 4 (10.7μm: window IR), and 5 (12μm: longwave IR). An arithmetic combination of these, which utilized the differences between brightness temperatures (particularly between Channels 4 and 5), yielded a unitless brightness count: $B = 60 + 10(T_5-T_4) + (T_2-T_4)$. Values of B that were large in comparison to the surrounding clouds and terrain represent volcanic ash (Ellrod, 98). The same is true for the GOES-12 volcanic ash algorithm, which also utilizes brightness temperatures in an arithmetic combination. However, brightness temperatures are taken from Channels 2, 4, and 6 (13.3μm: longwave IR): $B = 5(T_2 - 1.5T_4 + 1.5T_6) - 230$ (Ellrod, 04a). Different brightness count values represent different shades of gray for each algorithm. These shades of gray comprise an IR image.

There have been concerns that the GOES-12 volcanic ash algorithm, utilizing the 13.3μm channel in the absence of the 12μm channel, will result in a degradation of the display of volcanic ash data. The 12μm channel had been used successfully in a two-band "split window" technique for about a decade to detect volcanic ash (Ellrod, 04b). An impact study has shown that some degradation is likely in dust, smoke, and volcanic ash detection, as well as detection of low cloud. Several studies comparing IR data from GOES-11 and GOES-12 have already demonstrated this for the Aleutian Island chain (Schmit, 01) and for Popocatepetl (Ellrod, 01) volcanoes. Volcanic ash detection will especially be degraded in the presence of significant amounts of high cirrus clouds or diffuse ash. These degradations are also expected to be more pronounced during the night (Schmit, 01).

The reason why accurate volcanic ash detection is so crucial is that volcanic ash clouds, if undetected, can be dangerous and even fatal for planes. If the ash is ingested into the jet engines, volcanic ash decreases their efficiency. Furthermore, the plane's leading edge surfaces can be damaged (Ellrod, 98). Recently, the increased amount of air traffic in the circumpacific region of the globe where the most potentially active volcanoes are located has increased the necessity for accurate volcanic ash data. Modern jets also have higher temperatures that are within the melting range of volcanic glass in the ash. In particular, modern jet engines are quite susceptible to ash damage during the take-off and landing periods (Kite-Powell, 01). Although few studies have yet been done on the effect of the 12μm band on the detection of cloud data, this is essential for monitoring environmental conditions and for forecasting purposes, especially for convective clouds and possibly dangerous conditions for planes and those on the ground.

The purpose of this project is to compare the capabilities of the GOES-11 volcanic ash algorithm with the 12μm band to the GOES-12 volcanic ash algorithm with the 13.3μm band for detecting volcanic ash and clouds. There is a GOES-11 (GOES-West) satellite still in orbit over the Pacific Ocean. This and the GOES-12 (GOES-East) satellite have an area of overlap in Mexico and Central America, making it possible to compare data from two active volcanoes in Mexico: Colima and Popocatepetl. The expected result is that the GOES-11 volcanic ash algorithm will provide better volcanic ash data and cloud data than GOES-12.

2. Methods and Materials

Data for the volcanoes of Colima and Popocatepetl during July and August 2006 was generated on an archive (see ftp.orbit.nesdis.noaa.gov/pub/smcd/opdb/ematson) using the program McIDAS (Man computer Interactive Data Access System). This data included IR images using volcanic ash algorithms for the GOES-11 Imager and the GOES-12 Imager. The IR images were further divided into images without any contour, images overlaid with a brightness temperature contour from Channel 4, and images overlaid with a brightness count contour. In addition, visible (VIS) images (0.6μm: Channel 1) were generated from GOES-12, and scatter plots of brightness temperature vs. brightness count were generated for both GOES-11 and GOES-12.

The VIS images proved to be extremely important as an accurate source of information to compare the IR data to. After a VIS image from a particular day at a particular time was examined, it was compared to the IR images. This was a good test to see which volcanic ash algorithm provided more accurate data. The scatter plots were also consulted to determine the correlation between brightness temperature and brightness count at various times of day for each satellite. Various volcanic watch websites, including the Washington VAAC and el Monitoreo Volcánico in Mexico, were also consulted. The Colima data was consulted more frequently than the Popocatepetl data, as there was only one IR image each day generated by GOES-11 for Popocatepetl.

Since the official time as measured from the satellites was in Greenwich Mean Time (GMT) or Coordinated Universal Time (UTC), 5 hours were subtracted from the recorded time when the local time (LT) in Popocatepetl or Colima was desired. The data collected around midday was particularly important since, at that time, the sun angle would be nearly the same for both satellites and there would be the least difference in the measured heat (IR) and reflectivity (VIS). Since the subpoint of GOES-11 was a bit further away from Colima (by about 1000 miles) and Popocatepetl than GOES-12, displacement, which is a result of the viewing angle of the satellite, was also noted. Due to displacement, the thicker and convective clouds were displayed a little too far to the east in the GOES-11 images. Other errors that were taken into consideration were attenuation, lack of background contrast, and contamination. Limitations, including time and the available archives, were also taken into account.

3. Case Studies

*a. Colima Images from 27 July 2006 18:00 UTC*

The images in Figure 1 were all archived at 18:00 UTC, or 13:00 LT, around early afternoon. In the VIS image, there is a very thick, pronounced band of cirrus clouds in the center with a thin, translucent cirrus band to the east and some thinner, middle cloud layers to the west. In the GOES-11 IR image, the thin cirrus band is distinguishable from the thick cirrus. However, in the GOES-12 IR image, it is not. Neither satellite was able to detect the small cumulus clouds to the east. There is a greater contrast between the thick cirrus and middle cloud layer in the GOES-12 imagery. However, there is not as great of a contrast even in the VIS imagery.

Concerning thick cloud layers, it was found that GOES-11 provided a better contrast between the higher and lower layers than GOES-12. It provided better coverage for thin cloud edges than GOES-12, which often did not detect these edges.

*b. Colima Images from 3 July, 2006 0:00 UTC*

The images in Figure 2 were archived at about 0:00 UTC or 19:00 LT, in the evening. The VIS image shows some very distinct overshooting tops over Colima and to the west as part of a convective storm complex. There are also lower clouds to the east and over the volcano of Paricutin. These overshooting tops are highlighted quite well in the GOES-11 IR image. Their cold centers are also clearly shown. However, in the GOES-12 imagery, they are hardly distinguishable from the surrounding terrain. There is little, if any, evidence of cold centers. The lower clouds over Paricutin were detected by GOES-11, but not detected by GOES-12.

Concerning cumulonimbus clouds, GOES-11 in general was found to be more effective in detecting overshooting tops than GOES-12. However, certain cumulonimbus clouds without overshooting tops were detected with roughly the same accuracy and brightness values by GOES-11 and GOES-12. The same was true for mid-level clouds and cumulus clouds (which were detected equally poorly by both).

*c. Colima Scatter Plots from July 19, 2006*

Examining the brightness temperature vs. brightness count scatter plots proved to be extremely useful in determining patterns and data correlations (see Figure 3). For GOES-11, the general correlation was inverse proportionality (higher brightness counts for lower brightness temperatures). It was the opposite for GOES-12: direct proportionality (higher brightness counts for higher brightness temperatures). It was found that, in general, there was a better data correlation for GOES-11 than GOES-12. The worst correlation for GOES-11 was around 11:00 UTC, or 6:00 LT, in the early morning, and around 0:00 UTC, or 19:00 LT, in the evening, for GOES-12. These differences were attributed to the varying sun and satellite sensor viewing angles.

The best correlations for GOES-11 were in the afternoon, around 20:00 UTC, or 15:00 LT. Quite surprisingly, the best correlations for GOES-12 seemed to be at night, around 5:00 UTC, or 0:00 LT. In the impact study done for the effect of the loss of the 12μm band, the degradations were expected to be more pronounced at night (Schmit, 01). After examining the IR images at night for both GOES-11 and GOES-12 (see Figure 4), it was found that in overcast places, there was an absence of data for GOES-11. The opposite was true for GOES-12: in overcast places, there was an excess of data and too much noise.

*d. Colima Scatter Plot Spikes from July 30, 2006*

On a more or less day-to-day basis, a strange spike was observed in the GOES-12 brightness temperature vs. brightness count scatter plots for Colima (see Figure 5). This spike occurred along the x-axis and at lower temperatures. However, it did not occur at low enough temperatures for it to have been caused by ground features. The time of day it occurred was around 22:00 UTC, or 17:00 LT: late afternoon. After estimating the sun angle at the time for GOES-12, it was speculated that the cause was sunglint off the Pacific Ocean. Although full disk IR images were not available to spot the sunglint, several GOES-12 VIS images were examined. The ocean was significantly brighter at the times of the spike.

There was also a similar spike observed on the GOES-11 scatter plots. However, it was observed a little earlier in the day, around 17:00 UTC, or 12:00 LT. This made perfect sense since the sun angles for GOES-11 and GOES-12 were usually different. Keeping this in mind, the time at which GOES-11 had the optimum sun angle that would cause the brightest sunglint would be different for GOES-12. Unfortunately, there was no VIS data from GOES-11 to examine. However, it made sense that the earlier GOES-11 spike was also due to sunglint on the Pacific Ocean.

## 4. Discussion and Conclusion

In interpreting these results, it was crucial that the possible errors be taken into account. Since the subpoint of GOES-11 was about 1000 miles further away from Colima than GOES-12, one recurring error was displacement. Especially with tall convective clouds and overshooting tops, the clouds in the GOES-11 IR imagery were displaced a little to the east. This difference in distance also caused the viewing angle of the satellite sensors to be different as well as the sun angle: considerations that made data from around midday especially important. Another error in the IR imagery was a lack of temperature contrast between the ground and low clouds, often making them difficult or impossible to distinguish. Attenuation, which causes clouds to appear cooler and higher in the atmosphere as a result of absorption and scattering of radiation, and contamination, which occurs when radiation reaches the sensor from beneath a cloud, were also taken into account.

Just as expected in the impact study (Schmit, 01), GOES-11 was shown to detect low clouds much better than GOES-12. The composition of low clouds is mostly water vapor. Since their temperatures are so close to that of the ground, they are often difficult to distinguish in IR imagery. This suggests that the contrast between the brightness temperatures detected by Channels 4 and 5 (the 12µm channel), or the "split-window" method, something that is not present in the GOES-12 volcanic ash algorithm, is important in detecting low clouds.

Other observations were not as expected. For instance, even though the "split-window" method was noted to be useful in the detection of thin cirrus clouds, the 13.3µm band was also noted useful for cirrus clouds (Schmit, 01). However, GOES-11 was shown to detect cirrus, especially thin cirrus, much better than GOES-12. Cirrus clouds are mainly composed of ice, and often have a wispy, wind-blown appearance. In addition to proving the usefulness of the "split-window" method in detecting thin cirrus, this suggests that contamination, caused by radiation from underneath a cloud, is a greater problem with GOES-12 than with GOES-11.

GOES-11 was also shown to contrast thick cloud layers and detect thin cloud edges with more clarity than GOES-12. However, concerning cumulus clouds, GOES-11 and GOES-12 provided equally poor data. Since cumulus clouds, mainly composed of water droplets, are irregularly shaped and quite small, this was due to the fact that neither satellite had good enough resolution to detect such small clouds. Mid-level clouds were also detected about equally and with similar brightness counts. Since, in general, mid-level clouds do not have temperatures that are close to those of the ground, the "split-window" method is not as important. The scatter plot data supported this finding for mid-level clouds. Since the data trends for GOES-11 and GOES-12 were opposite, with contrasting brightness values for warm and cold brightness temperatures, it made sense that the brightness counts for the medium temperature, mid-level clouds would be about the same.

For cumulonimbus and overshooting tops, on the other hand, GOES-11 was far superior. Not only were these clouds highlighted well, but so were their cold centers. GOES-12 highlighted overshooting tops poorly and did not show the cold centers at all. Cumulonimbus clouds are a mixture of water and ice, and develop overshooting tops when they are tall enough to break through the cirrus layer. Again, the superior detection of overshooting tops by GOES-11 shows the value of the "split-window" method in contrasting temperatures: in this case, for contrasting the cirrus and the overshooting tops.

The scatter plot data of brightness temperature vs. brightness count was also quite helpful in comparing the GOES-11 algorithm to GOES-12. The general trend for the GOES-11 data was inverse proportionality (higher brightness counts for lower brightness temperatures), and for GOES-12, direct proportionality (higher brightness counts for higher brightness temperatures). It was found that, in general, the GOES-11 scatter plots had better data correlations than GOES-12. The best and worst correlations for GOES-11 and GOES-12 occurred at different times of day due to the differences in sun angle. This difference in sun angle was also apparent in the curious spikes in the Colima scatter plots, speculated to be caused by sunglint from the Pacific Ocean.

Curiously, the best data correlation for GOES-12 occurred at night. Upon examination of the IR images at night, the overcast areas in the GOES-12 images lacked data values. The reverse was true for the GOES-11 images: in the overcast areas, there was too much noise, probably caused by the shortwave IR Channel 2. Since Channel 2 is a shortwave IR channel, both VIS and IR radiation are used to determine brightness temperature values. At night, in the absence of VIS radiation, this would cause the brightness temperature values to vary greatly. It is not known why the correlation for GOES-12 improved at night or why there was an absence of data values in overcast areas. Perhaps the fact that there were so few data values was what improved the correlation. Overall, the pronounced degradations at night that were predicted by the impact study were shown by the data (Schmit, 01).

Unfortunately, there was no volcanic ash data available for comparison due to poor satellite data and little volcanic activity. Although there was a prominent ash cloud emitted from Popocatepetl on 4 August, this ash cloud was not identifiable from either GOES-11 or GOES-12 satellite data. Since there was thick cloud cover that day, this emphasizes the need to improve the volcanic ash algorithm to distinguish ash from cloud cover, particularly cirrus (Schmit, 01).

In general, the expected degradations in cloud data (Ellrod, 04b) were observed in the GOES-11 and GOES-12 IR data. However, new observations were made that had not been researched in great detail in the past. GOES-11 was shown to detect cirrus (especially thin), cumulonimbus with overshooting tops, thick cloud layers, thin cloud edges, and low clouds better than GOES-12, and mid-level clouds and cumulus about the same. The scatter plot data emphasized the fact that the GOES-11 data was more accurate by showing much better data correlations. Limitations in this study included time, available archives, and the difference in distances of the sensors from Colima and Popocatepetl. Overall, the collected data indicated that the GOES-11 volcanic ash algorithm with the 12μm band and the "split-window" method provides much better data for various types of clouds. It emphasized the importance of the "split-window" method in providing contrasting temperatures between different surfaces. The importance of the 12μm band was highlighted, and scope for future research in this area was provided. This study will be extremely useful in deciding which channels to include in future GOES satellites: including GOES-R in the next decade (Ellrod, 04b).

## 6. Acknowledgements

The author would like to thank several people for making this project possible. First of all, to her mentor, Kenneth L. Pryor, for helping her in her project choice, instructing her in the basics of remote sensing, allowing her access to the data, and providing constant guidance in her research. Secondly, to Mr. Pearce for his hard work in arranging the mentorship program guidelines and the mentorship. Thirdly, to Mrs. Wu, the research advisor, for keeping track of the author's progress and being her sponsor for the science fair. The author would also like to acknowledge the work of Gary Ellrod, whose in-depth research on the volcanic ash product was extremely helpful.

## 7. Appendix

**Attenuation** – "A decrease in the amount of radiation reaching the satellite due to absorption or scattering by the intervening medium (Kidder, 95)." This error increases with the distance from the satellite subpoint, is most pronounced in IR bands, and causes clouds to look cooler and higher in the atmosphere.

**Blackbody** – "A theoretical object that absorbs 100% of the radiation that hits it (from www.daviddarling.com)." This means that there is no reflection, and the object appears totally black. Since a blackbody can absorb any wavelength of energy, this means it can also emit any wavelength. As temperature increases, the peak wavelength (where the most radiant energy is emitted) emitted by the blackbody decreases.

**Brightness temperature** - "The temperature a blackbody would need to emit radiation at the observed intensity (radiance) at a certain wavelength (from www.daviddarling.com)." Brightness temperatures are not measured directly, but rather derived from inverting the Planck function after values are plugged in for radiance and wavelength. This is done for each pixel, whose size is determined by the resolution of the channel.

**Channel -** A 'band,' or 'channel,' is a range of wavelengths, centered on a particular wavelength, for which the satellite measures electromagnetic energy. All other wavelengths are filtered out. This is because there are only certain wavelengths that can penetrate the atmosphere and define the Earth's surface features from a satellite, located in what are known as 'atmospheric windows.'

Channels for GOES-11 and GOES-12:

|      | **GOES 8-11**       |                 | **GOES 12**         |                 |
| ---- | ------------------- | --------------- | ------------------- | --------------- |
| Band | Wavelength ($\mu m$) | Resolution (km) | Wavelength ($\mu m$) | Resolution (km) |
| 1    | 0.6                 | 1               | 0.6                 | 1               |
| 2    | 3.9                 | 4               | 3.9                 | 4               |
| 3    | 6.7                 | 8               | **6.5**             | **4**           |
| 4    | 10.7                | 4               | 10.7                | 4               |
| 5    | 12.0                | 4               | -                   | -               |
| 6    | -                   | -               | 13.3                | 8               |

**Contamination** – Caused by radiation being sensed through a cloud, this error is most pronounced with cirrus (Kidder, 95). This makes clouds appear warmer and closer to the ground.
**Cirrus** – Thin, wispy, ice crystal clouds at high altitudes (Conway, 97).
**Cumulonimbus** – Large cumulus clouds that can grow to 60,000 feet and are capable of producing extreme weather conditions such as thunderstorms, hail, heavy rain, and tornadoes (Conway, 97).
**Cumulus** – Low-altitude, fair weather clouds with limited vertical development and characterized by a popcorn or cotton ball shape (Conway, 97).
**Displacement** – An error that occurs as a result of the viewing angle geometry and the effect of a 3D image being projected onto a 2D surface, displacement increases with the distance from the satellite subpoint. This error is more pronounced with tall clouds and results in the tops of the clouds being displaced to a point away from the satellite subpoint (Kidder, 95).
**Geosynchronous satellite** – A satellite whose period of revolution is the same as the Earth's. This is what the GOES satellites are, allowing them to continuously hover over the same spot on the Earth's surface. This is a definite advantage for providing continuous data needed for concentrated data analysis (from www.oso.noaa.gov).
**Infrared (IR) radiation** – The portion of the electromagnetic spectrum that is felt as heat (Conway, 97): measured by satellite sensors as emitted thermal energy.

**LCT (Local Civil Time)** – The local time at a particular Earth location.
**Noise** – Random electrical signals, introduced by circuit components or natural disturbances, that tend to generate errors in transmission.
**Overshooting tops** – "A phenomenon that occurs when towering convective clouds break through the cirrus top of a thunderstorm and give the cloud top a small area of a very lumpy texture. Overshooting tops occur in regions where the updraft region of the storm is intense enough to break through the cloud top (Conway, 97)."
**Planck function** – An equation that determines the radiance being emitted by a blackbody at a specific wavelength and a specific temperature:
$$B_\lambda(T) = (c_1 \lambda^{-5})/(\exp(c_2/(\lambda T)) - 1)$$
Where $B_\lambda(T)$ is the radiance, $c_1$ and $c_2$ are radiation constants, T is the temperature, and $\lambda$ is the wavelength (Kidder, 95).
**Radiance** – Radiance flux density per unit solid angle, or radiant energy per unit time crossing a unit area per unit solid angle (Kidder, 95).
**Resolution** – The size of the smallest element in the viewing field that can be detected by the satellite (Kidder, 95). The resolution is determined by the area of the smallest pixel, or point on the viewing grid of the satellite.
**Stratus** – Low-level clouds, including fog, with a flat appearance (Conway, 97).
**Subpoint** – The point on the Earth that is directly beneath the satellite. There is the best resolution here because, since the Earth's curvature is not an issue, a smaller area is being viewed (Kidder, 95).
**Sun angle** – The zenith angle of the sun, measured from the vertical, at which the sun's rays are reflected off the Earth's surface and into the satellite sensor.
**Sunglint** – The reflection of the sun off of the water surface into the satellite sensor. It occurs when the satellite's viewing angle equals the angle of the sun's rays. This is particularly useful in wind detection since there is very little sunglint in rough areas. However, with calm waters, there is very bright sunlight in a localized area. With the GOES satellites, the sunglint moves from east to west each day as the solar subpoint changes (Conway, 97).
**UTC (Coordinated Universal Time)** –Refers to the time in Greenwich, England, which serves as the basis for legal civil time all over the Earth. It is also referred to as Zulu time (Z), or GMT (Greenwich Meridian Time).
**Viewing angle** – Caused by increasingly large zenith angles as a satellite gathers data from areas farther away from its subpoint, this error causes the satellite to look at the sides of clouds in addition to the tops. This causes the apparent portion of the sky that is covered with clouds to increase.
**Visible (VIS) radiation** – The form of electromagnetic energy that human eyes detect (Conway, 97): measured from satellite sensors as reflected solar energy.

8. Figures

a)

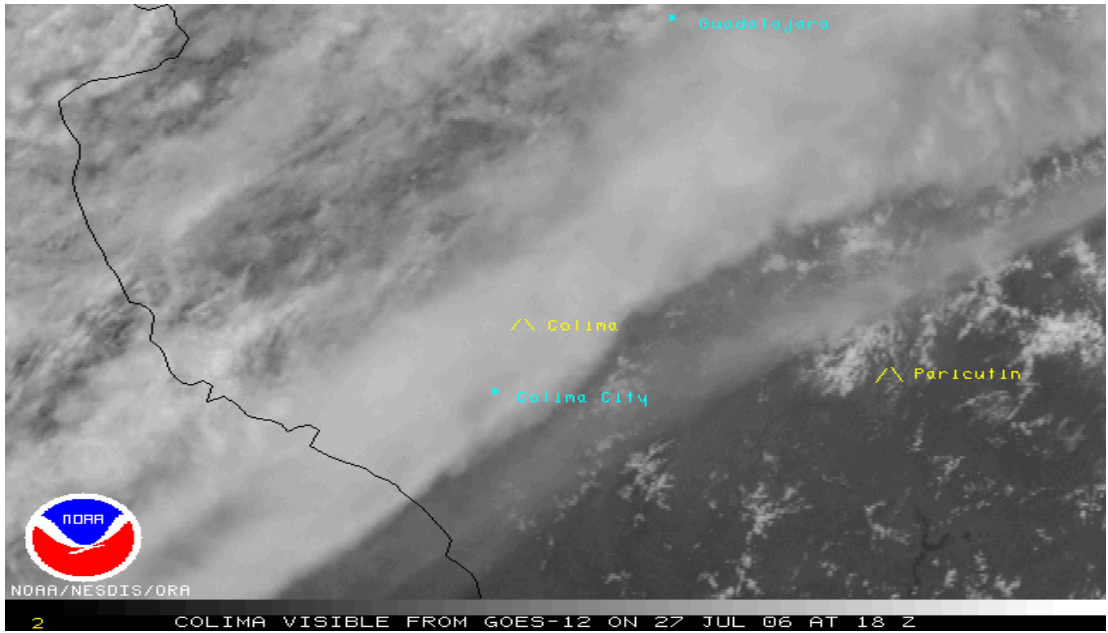

b)

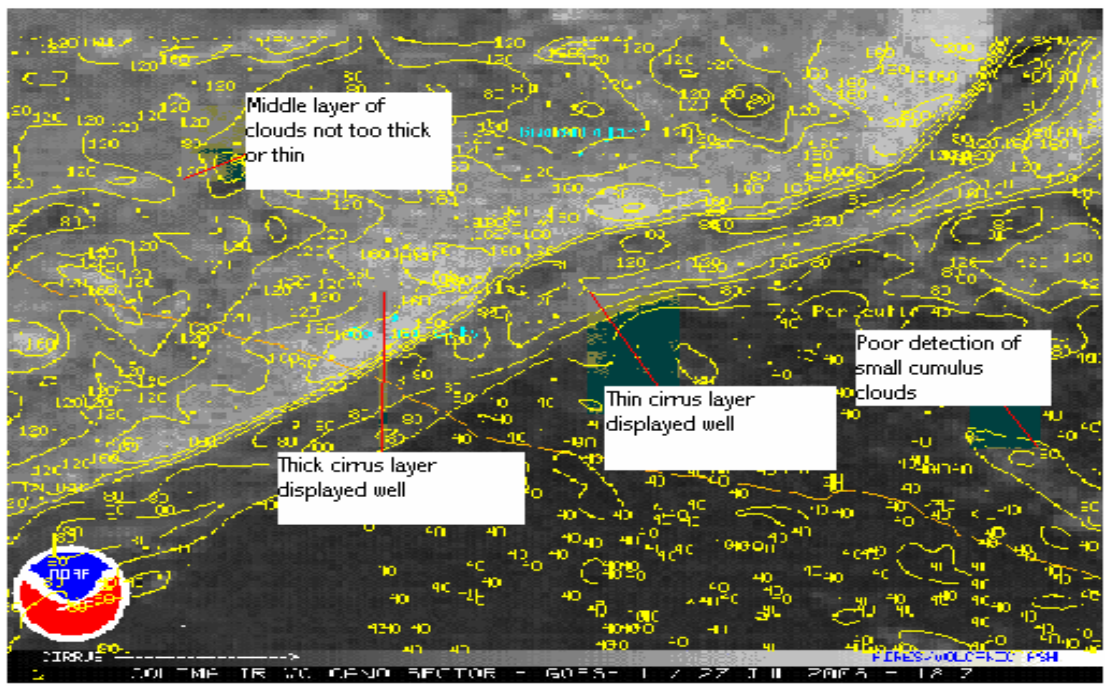

c)

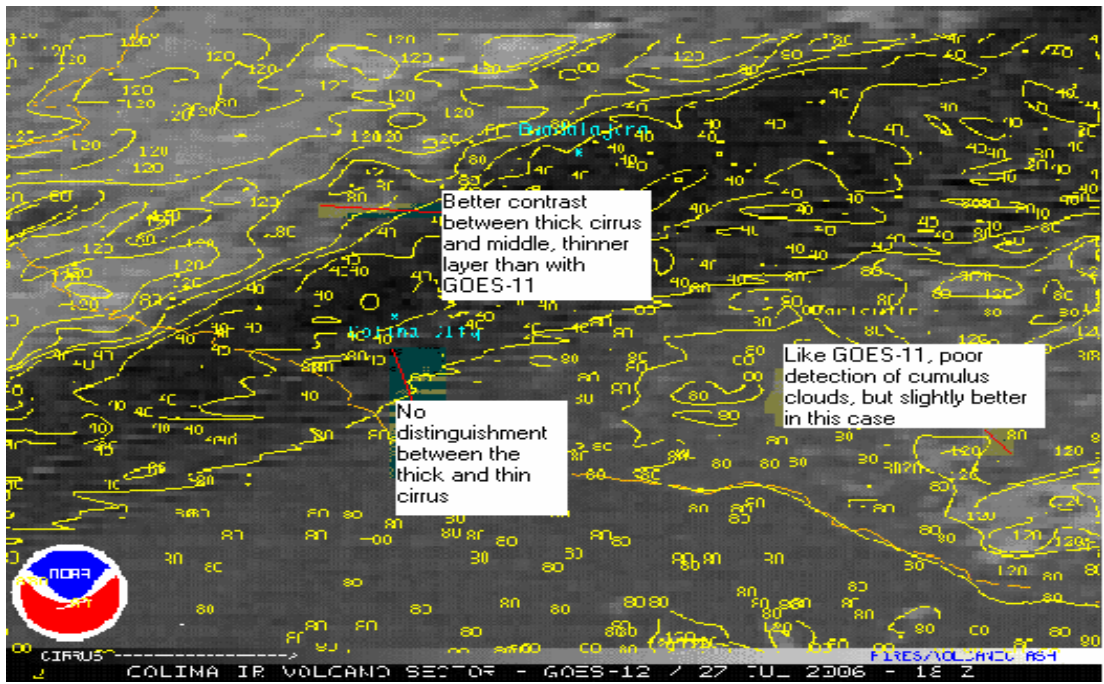

Figure 1. GOES imagery for Colima 27 July 2006 18:00 UTC a) GOES-12 VIS; b) GOES-11 IR (with brightness count contour); c) GOES-12 IR

a)

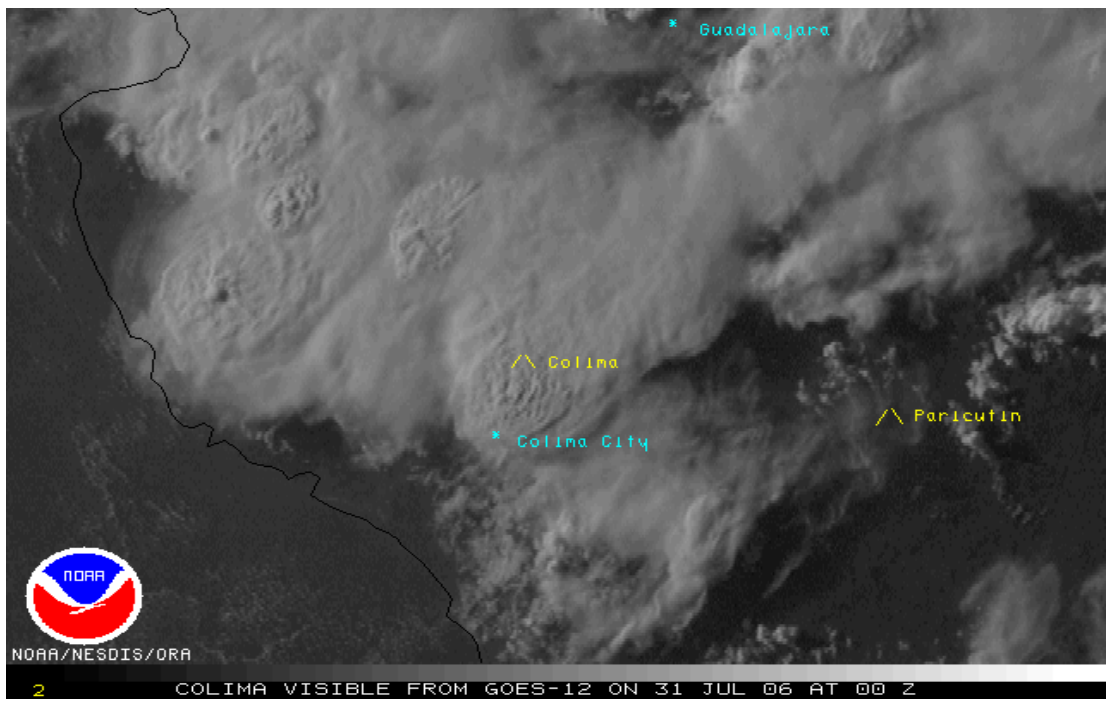

b)

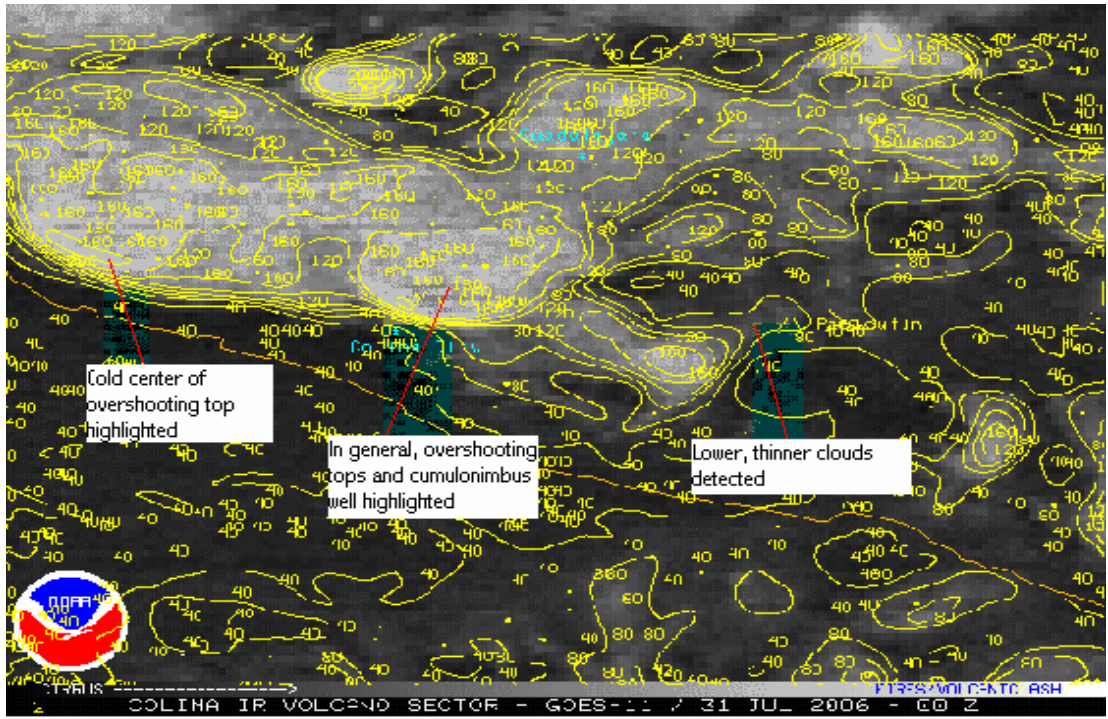

c)

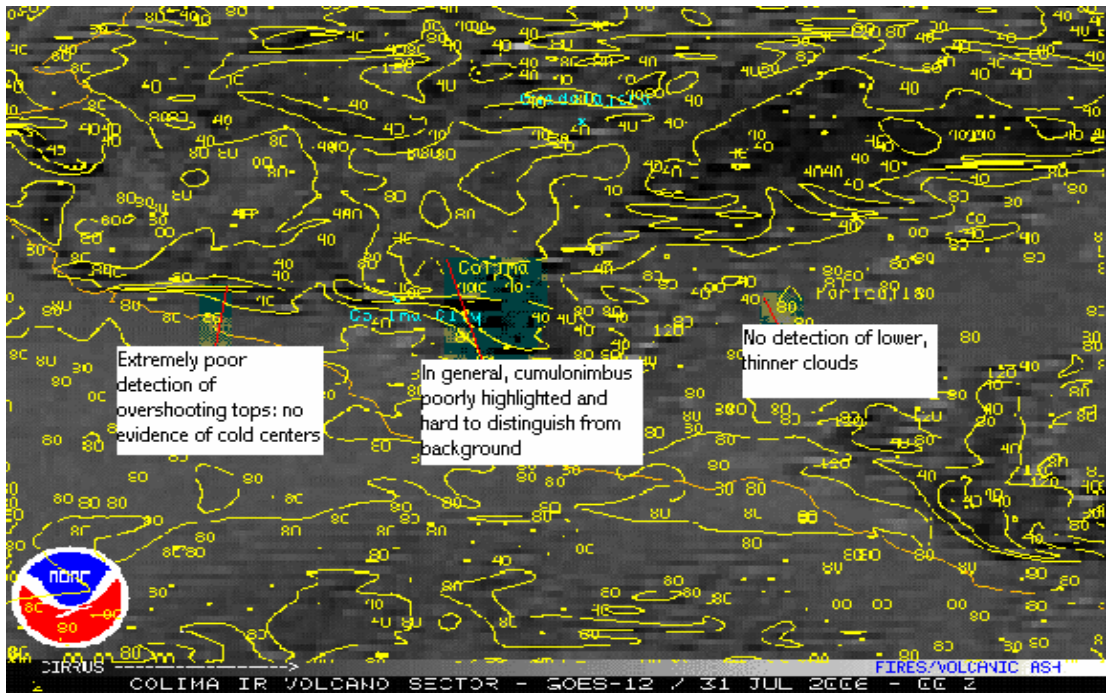

Figure 2. GOES Imagery for Colima 31 July 2006 0:00 UTC a) GOES-12 VIS; b) GOES-11 IR; c) GOES-12 IR

a)

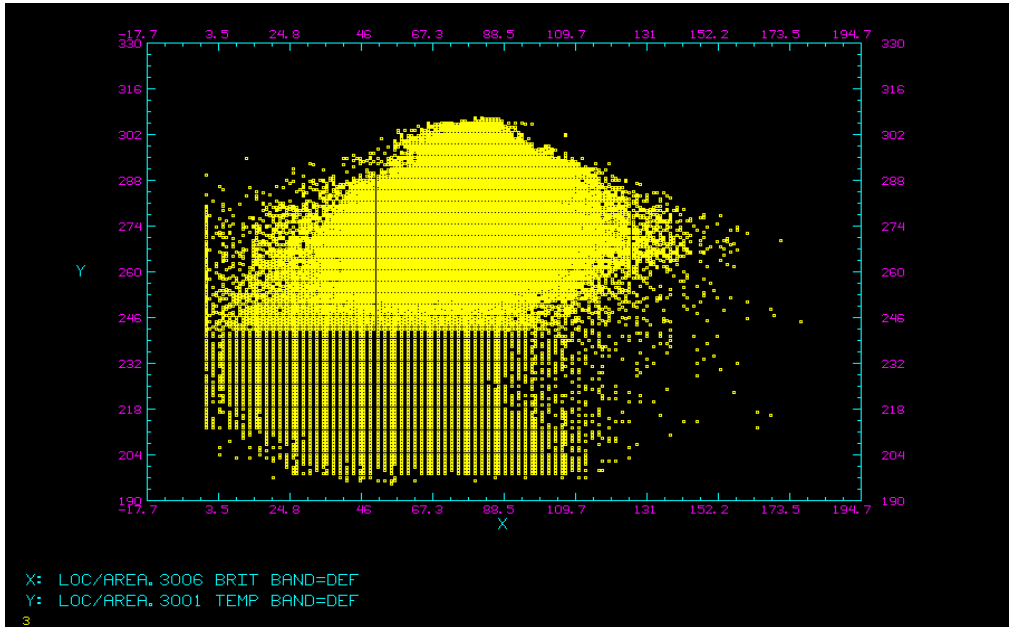

b)

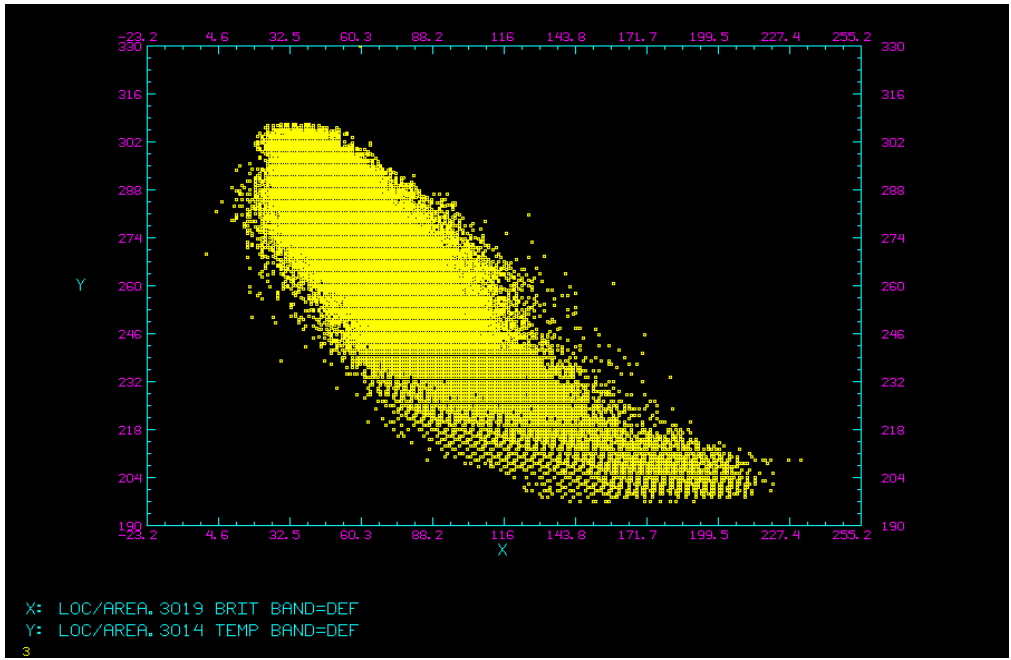

c)

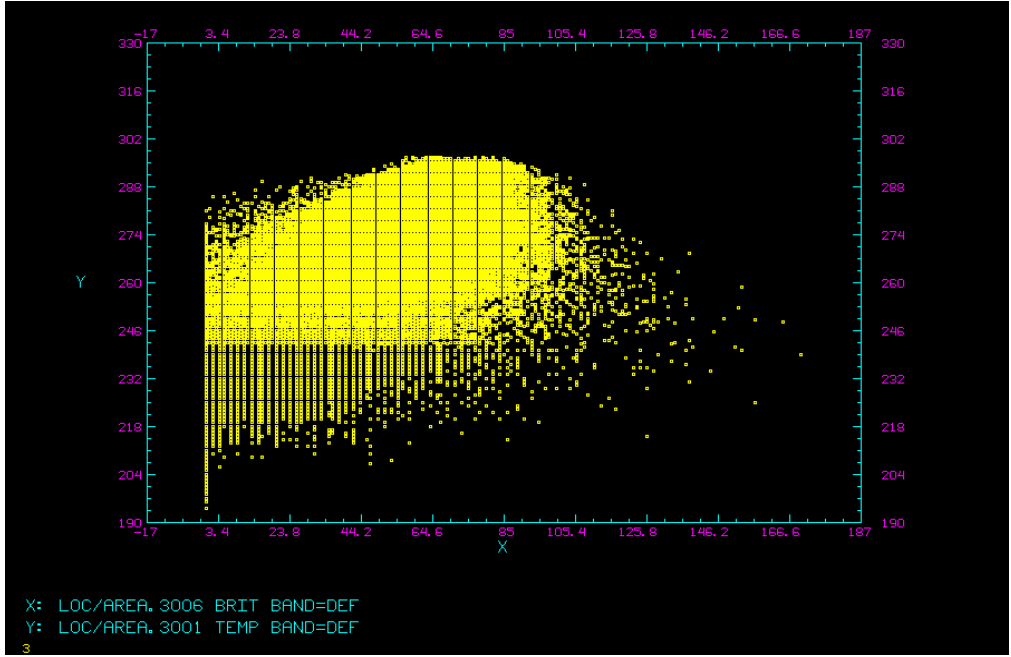

d)

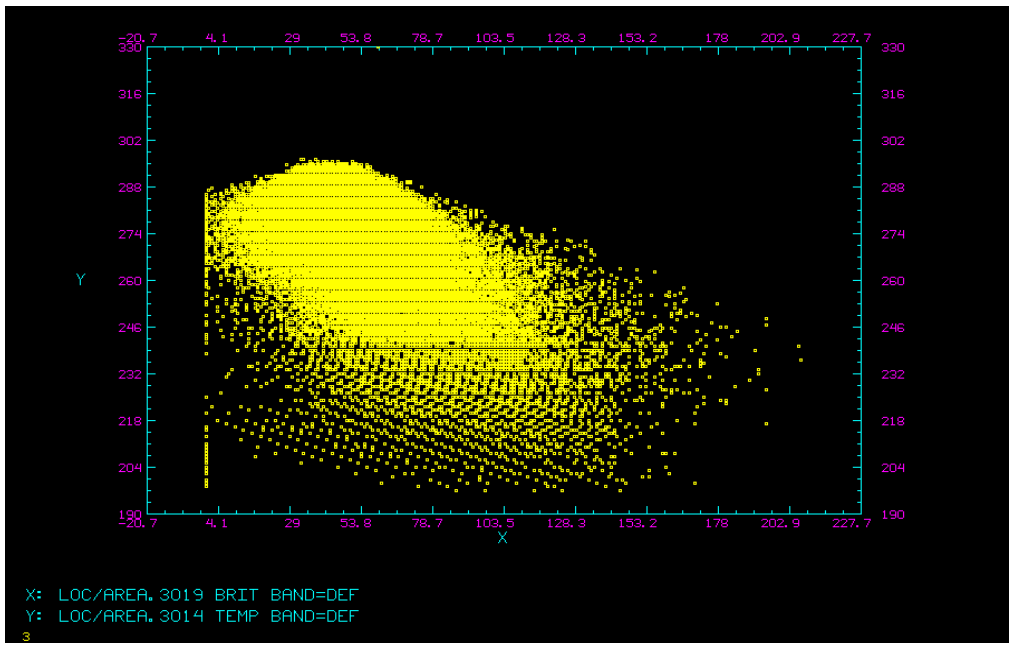

Figure 3. GOES brightness temp. vs. brightness count scatter plots for Colima a) GOES-12 19 July 2006 0:00 UTC; b) GOES-11 19 July 2006 0:00 UTC; c) GOES-12 19 July 2006 10:00 UTC; d) GOES-11 19 July 2006 11:00 UTC

a)

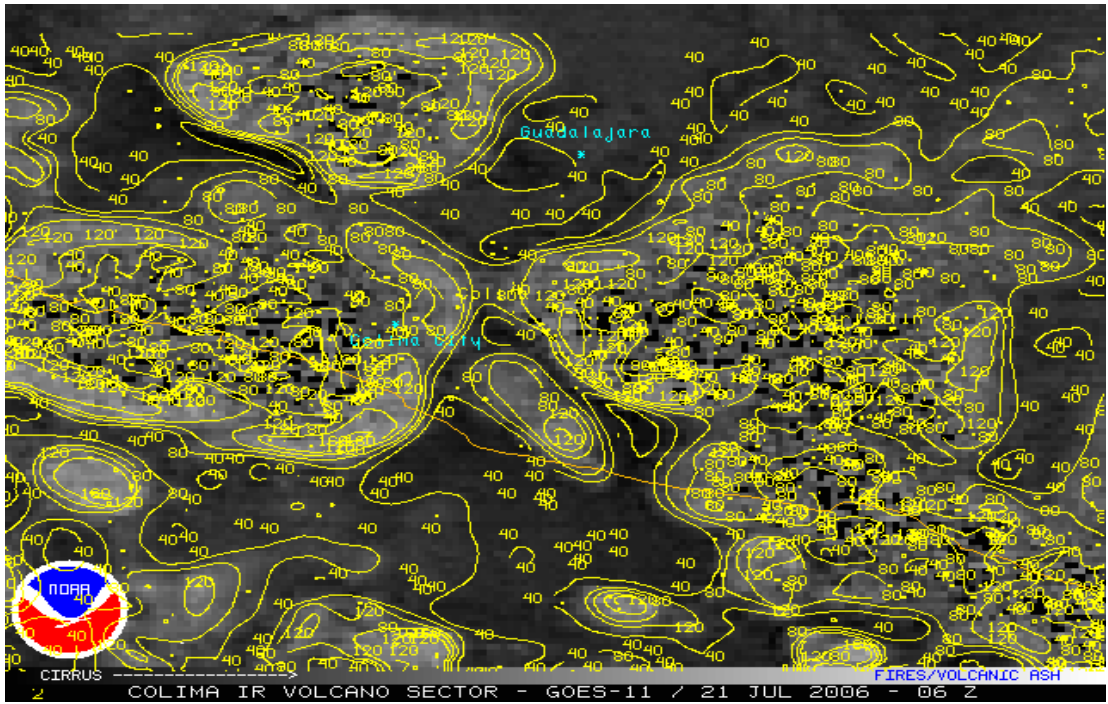

b)

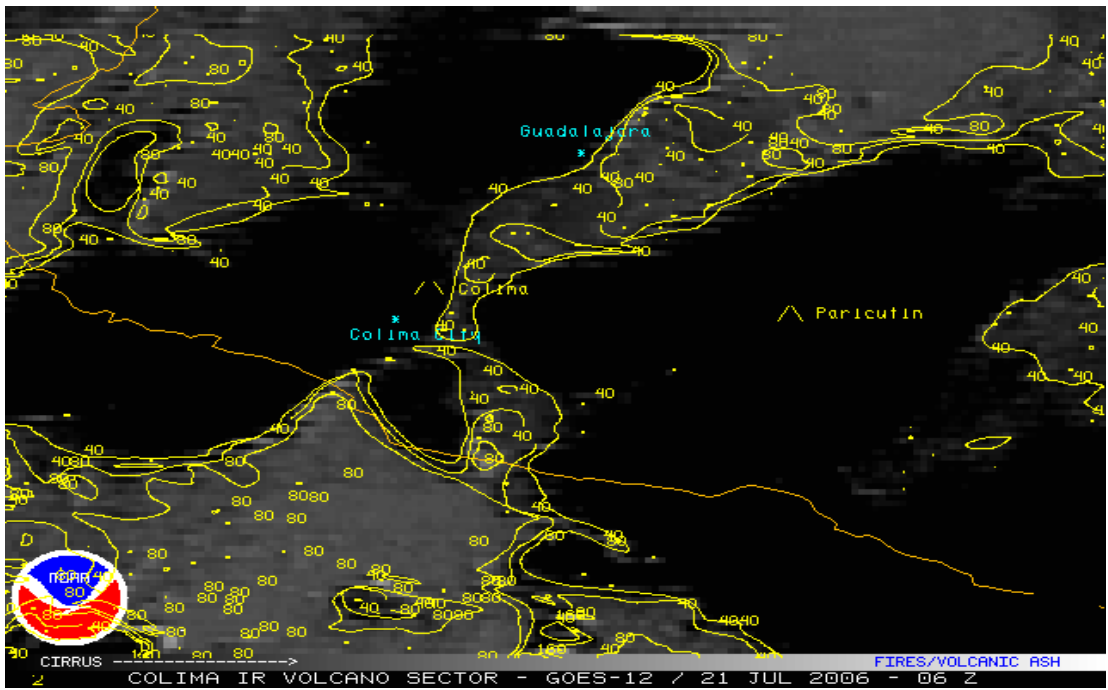

Figure 4. GOES IR imagery for Colima 21 July 2006 6:00 UTC for Colima a) GOES-11; b) GOES-12

a)

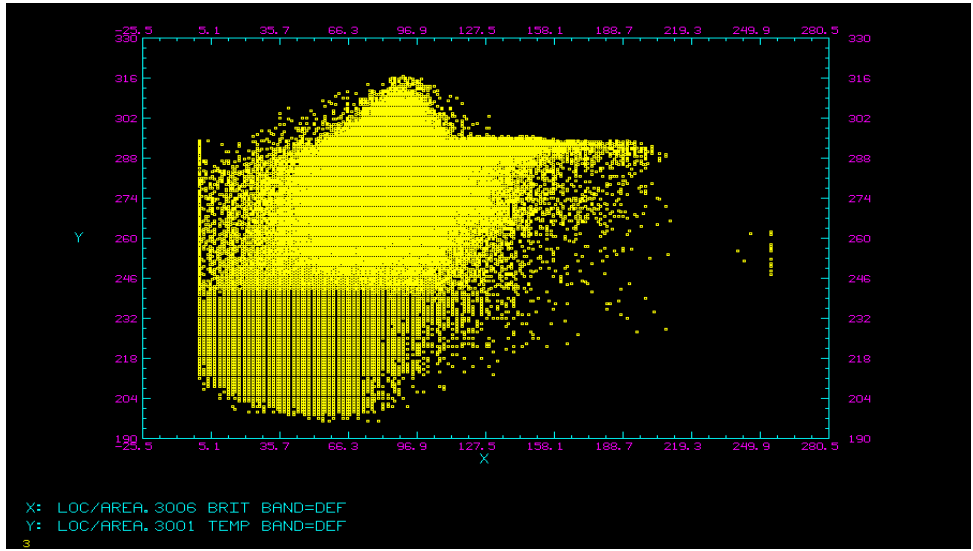

b)

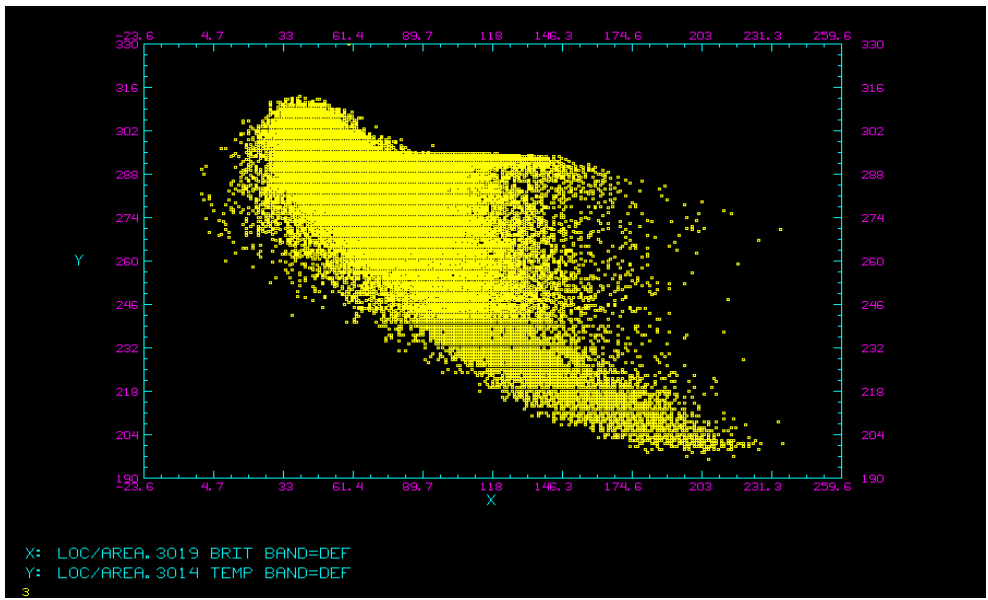

Figure 5. GOES brightness temp. vs. brightness count scatter plots 30 July 2006 for Colima a) GOES-12 22:00 UTC; b) GOES-11 17:00 UTC